\documentclass
[aps,prd,twocolumn,superscriptaddress,amsfont,graphicx,nofootinbib]{revtex4}%
\usepackage{color,graphicx,epsfig}
\usepackage{ifpdf}
\usepackage{amsmath}
\usepackage{bm}
\usepackage{color}
\usepackage[english]{babel}
\usepackage{graphicx}%
\usepackage{amsfonts}%
\usepackage{amssymb}
\usepackage{braket}
\usepackage{hyperref}
\usepackage{ifpdf}

\definecolor{nicered}{rgb}{0.7,0.1,0.1}
\definecolor{nicegreen}{rgb}{0.1,0.5,0.1}
\hypersetup{colorlinks,citecolor= nicegreen,linkcolor= nicered}

\newcommand{\slashed}{\slash \hspace{-0.19cm}}

\newcommand{\op}{{\cal O}}
\newcommand{\J}{{\cal J}}
\newcommand{\DM}{{\rm DM}}

\newcommand{\bea}{\begin{eqnarray}}
\newcommand{\eea}{\end{eqnarray}}
\newcommand{\beq}{\begin{equation}}
\newcommand{\eeq}{\end{equation}}

\newcommand{\diag}{\operatorname{diag}}

\begin{document}

\title{Discovering Dark Matter Through Flavor Violation at the LHC}

\def\LjubljanaFMF{Faculty of Mathematics and Physics, University of Ljubljana,
 Jadranska 19, 1000 Ljubljana, Slovenia }
\def\Cincy{Department of Physics, University of Cincinnati, Cincinnati, Ohio 45221,USA}
\def\LjubljanaIJS{Josef Stefan Institute, Jamova 39, 1000 Ljubljana, Slovenia}

\author{Jernej F. Kamenik}
\email[Electronic address:]{jernej.kamenik@ijs.si} 
\affiliation{\LjubljanaIJS}
\affiliation{\LjubljanaFMF}

\author{Jure Zupan} 
\email[Electronic address:]{jure.zupan@cern.ch} 
\affiliation{\LjubljanaIJS}
\affiliation{\LjubljanaFMF}
\affiliation{\Cincy}

\date{\today}

\begin{abstract}
We show that the discovery channel for dark matter (DM) production at colliders can be through flavor violating interactions resulting in a novel signature of a single top and large missing transverse energy. We discuss several examples where the production of DM is dominated by flavor violating couplings: minimal flavor violating models with a large bottom Yukawa, models with horizontal symmetries, and DM  in nontrivial flavor group representations. Discovery at the 7 TeV LHC with a few fb${}^{-1}$ may already be possible. 
\end{abstract}

\maketitle

{\bf Introduction.}
The matter fields of the Standard Model (SM) come in three generations, leading to distinct  flavors of quarks and leptons. The gauge interactions do not distinguish between different generations and are flavor blind. The Yukawa interactions, on the other hand, are flavor violating. We focus on the quark sector, where the eigenvalues of the Yukawa matrices - the quark masses - are very hierarchical and span 5 orders of magnitude. Similar hierarchical structure is seen in the Cabibbo-Kobayashi-Maskawa quark mixing matrix, where the smallest off-diagonal element is $V_{ub}\simeq 3\times 10^{-3}$.

A distinguishing feature of the SM gauge and matter structure is that no Flavor Changing Neutral Currents (FCNCs) are generated at the leading perturbative order. They are further suppressed also by the smallness of the relevant CKM matrix elements. The agreement of predicted small FCNCs with the precision flavor experiments requires any New Physics (NP) at the TeV scale to have a highly nontrivial flavor structure. Only small amount of flavor violation is allowed phenomenologically. The flavor violation cannot be completely absent, however. If nothing else, the flavor symmetry is broken already  by the SM Yukawas. At least at loop level (and thus also from RG running) these will then feed into the interactions between NP and the SM sector. Thus some amount of flavor violation in the interactions between NP and SM sector is unavoidable.

In this Letter we explore the consequences of the above insight for the detection of Dark Matter (DM) at colliders. We will show that large effects are likely, leading to a prominent signal of a single top plus missing transverse energy (MET). A $t+\slashed E_T$ final state is an experimentally readily accessible channel. Since in the SM the production is both loop and CKM suppressed an observation of a $t+\slashed E_T$ signal above the background would be a clear signal of NP at LHC. In fact, the $t+\slashed E_T$ could even be a discovery channel of DM for a large set of NP models. For instance, the cross section for $t+\slashed E_T$ can  be orders of magnitude larger then the monojet cross section even in the case of Minimal Flavor Violation (MFV), if the interactions are chirality flipping. Somewhat surprisingly, DM would then be discovered through flavor violating interactions. While this paper was being finalized an analysis of  $t+\slashed E_T$ experimental reach at LHC appeared in \cite{Andrea:2011ws}, where a name {\it monotop} was coined for the $t+\slashed E_T$ signature. 

{\bf Effective field theory description.}
We want to compare the flavor violating production of DM at colliders with the flavor conserving one. The comparison crucially depends on the size of flavor violation in the NP sector that contains DM. 
To start with let us  make  the discussion quite general by using the simplifying assumption that all the NP states apart from DM are heavy enough so that we can integrate them out at a large scale $\Lambda$ (we will later relax this assumption).  We can then write down an Effective Field Theory (EFT) for DM interactions with the SM quark matter sector 
\beq
{\cal L}_{\rm int}=\sum_a \frac{C_a}{\Lambda^{n_a}} \op_a\,.\label{sumC}
\eeq
The sum above runs over the full set of SU(2) gauge invariant operators $\mathcal O_a$ that are bilinear in quark fields. For simplicity we assume that DM is not charged under SM gauge group, so that to $\mathcal O(n_a\leq3)$
\begin{align}
\op_{1a}^{ij}=&\big(\bar Q_L^i \gamma_\mu Q^j_L\big)\J_a^\mu\,, & &\nonumber\\
\op_{2a}^{ij}=& \big(\bar u_R^i \gamma_\mu u^j_R \big) \J_a^\mu,  &\op_{3a}^{ij}=&\big(\bar d_R^i \gamma_\mu d^j_R \big)\J_a^\mu\,, \label{op4a} \\
\op_{4a}^{ij}=&\big(\bar Q_L^i H  u^j_R\big) \J_a\,,& \op_{5a}^{ij}=&\big( \bar Q_L^i \tilde H d^j_R\big) \J_a\,,  \nonumber
\end{align}
and we do not write down additional tensor operators (contractions of Lorentz tensors $\J_a^{\mu\nu}$) for which the same discussion as for $\op_{4a,5a}$ will apply.
Here $Q_L,u_R,d_R$ are respectively the left-handed quark doublets, and right-handed up- and down- quarks, $i,j$ are the generational indices, $H$ is the SM Higgs doublet (with $\tilde H=i\sigma_2 H^*$), while ${\mathcal J}_{a}$ are the DM currents. Throughout this paper we assume that DM is odd under an exact $Z_2$. For  fermionic DM $\chi$ we then have ${\mathcal J}_{V,A}^\mu=\bar \chi \gamma^\mu\{1,\gamma_5\} \chi$, ${\mathcal J}_{S,P}=\bar \chi\{1,\gamma_5\} \chi$,  
(for Majorana fermion ${\mathcal J}_{V}^\mu=
0$), leading to $n_a=2$ for $\op_{1a,\dots,3a}$  in Eq. \eqref{sumC}, while for $\op_{4a,5a}$ we have $n_a=3$.
For scalar DM  ${\mathcal J}=\chi^\dagger \chi$, ${\mathcal J}^\mu=\chi^\dagger \partial^\mu \chi$, so that $n_a=2$ for all operators in
\eqref{op4a}. 

If DM is light enough the above operators can lead to FCNC decays of top~\cite{Li:2011ja}, $b$~\cite{Bird:2004ts} and even lighter quarks~\cite{Smith:2010st}. The last two are bounded by searches for the $b\to s \nu \bar \nu$ and $s\to d \nu\bar\nu$ decays, $Br(B^+ \to K^+ \nu\bar \nu)< 1.4 \times 10^{-5}$ \cite{:2007zk}, $Br(B\to K^*\nu\bar \nu) < 8.0\times10^{-5}$ \cite{:2008fr} and $Br(K^+\to\pi^+\nu\bar\nu)=(1.73^{+1.15}_{-1.05})\times 10^{-10}$~\cite{Artamonov:2008qb}. The reach for $Br(t\to j+2\chi)$ at 14 TeV LHC was estimated in Ref. \cite{Li:2011ja} to be ${\mathcal O}(10^{-4})$ for $5\sigma$ discovery with $10$ fb${}^{-1}$.

There are contributions to $B_{d,s}-\bar B_{d,s}$ and $K-\bar K$ mixing with DM running in the loop and two insertions of operators $\op_{1a,3a,5a}$. This gives the following bounds for couplings to the third generation \cite{Bona:2007vi}
\beq
\frac{C_{1a}^{13}}{\Lambda}\lesssim \frac{1}{2~{\rm TeV}}, \quad \frac{C_{1a}^{23}}{\Lambda}\lesssim \frac{1}{0.3~{\rm TeV}},
\eeq
and bounds of similar size for $C_{3a,5a}^{13,23}$. The bounds on $C_{2a,4a}^{13,23}$ on the other hand, come from top decays and are so loose that the EFT description breaks down before they are saturated. This indicates that large $t+\slashed E_T$ production signals from flavor violating couplings are possible at LHC and Tevatron.  It would be interesting to see, whether the more constrained (and thus more likely to come from flavor conserving operators)  $b+\slashed E_T$ NP signal can be picked out from the SM background of (mistagged) jet+invisibly decaying $Z$ events. From now on we focus on the more promising  $t+\slashed E_T$ channel and estimate its size in a number of models of flavor. 

\begin{figure}
\includegraphics[width=0.4\textwidth]{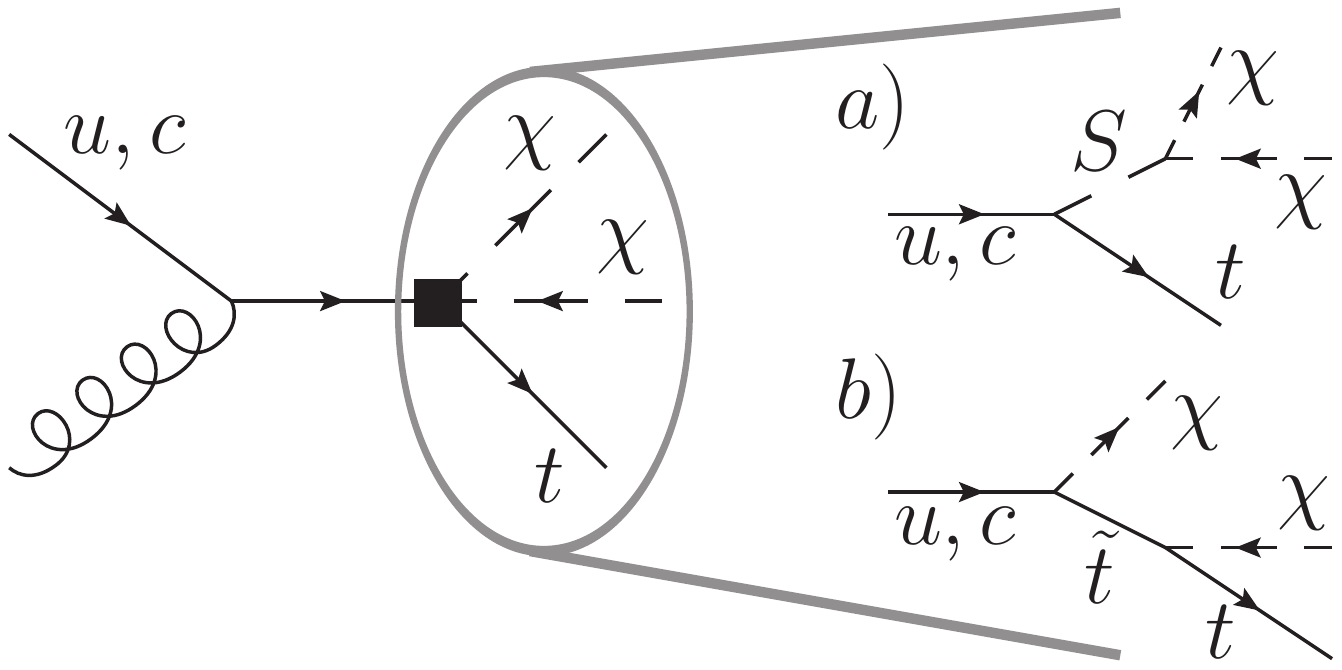}
\caption{{\footnotesize Flavor violating DM production at collider in the EFT description (left) and for two on-shell models, (a) with a SM gauge singlet $S$, and (b) with a color triplet  $\tilde t$ as a mediator.}}
\label{production:diagram}
\end{figure}

{\bf Minimal Flavor violation.}
Let us first assume that the interactions of the mediators with the SM are minimally flavor violating, i.e. that the flavor is only broken by the SM Yukawas $Y_{u,d}$. Using the spurion analysis \cite{D'Ambrosio:2002ex} the Wilson coefficients take the form
\begin{subequations}
\begin{align}
C_{2a}&=b_1^{(2a)}+b_2^{(2a)} Y_u^\dagger Y_u+b_3^{(2a)}Y_u^\dagger Y_dY_d^\dagger Y_u+\cdots,\\
C_{4a}&=\big(b_1^{(4a)}+b_2^{(4a)} Y_d Y_d^\dagger+\cdots\big)Y_u.
\end{align} 
\end{subequations}
In the up-quark mass eigenstate basis $Y_d=V_{\rm CKM}\diag(y_d,y_s,y_b)$ and $Y_u=\diag(y_u,y_c,y_t)$.  In the following let us assume that $b_1^{a}\sim b_2^{a}\sim b_3^{a}$ are all of the same order. The Wilson coefficient $C_{2a}$ is then flavor diagonal and universal to a good approximation and flavor violating interactions for all practical purposes are negligible. 

The situation is different for the chirality flipping operator $C_{4a}$ that is proportional to Yukawa matrix $Y_u$. In this case DM couples most strongly to the third generation, while the couplings to the first two generations are parametrically suppressed by $y_{u,c}/y_t$. 
This has important implications for the detection of DM at colliders. The flavor violating $q g\to t \chi \chi$ cross section is enhanced over the conserving  one by (see also Fig. \ref{production:diagram})
\beq
\begin{split}
\frac{\hat \sigma (u g\to t +2\chi)}{\hat \sigma (u g\to u+2\chi)}&\sim \left(\frac{y_t |V_{ub}| y_b^2}{y_u}\right)^2\sim 5\cdot 10^5 \,y_b^4, \\
\frac{\hat \sigma (c g\to t +2\chi)}{\hat \sigma (c g\to c +2\chi)}&\sim \left(\frac{y_t |V_{cb}| y_b^2}{y_c}\right)^2\sim 50 \, y_b^4.\label{MFVestimates}
\end{split}
\eeq
The $t+\slashed E_T$ signal can be significantly enhanced over the monojet signal even in the case of MFV, if two conditions are fulfilled, i) bottom Yukawa is large, preferably $y_b\sim O(1)$, and ii) DM couples to quarks through scalar interactions. We note in passing that DM coupling only through the SM Higgs portal would not lead to flavor violating effects. The above MFV counting thus assumes additional scalar interactions. Such interactions are for instance needed for isospin violating models proposed to explain CoGeNT and DAMA excesses \cite{Chang:2010yk} (see, however, also  \cite{Schwetz:2011xm}).

\begin{figure}
\includegraphics[width=0.45\textwidth]{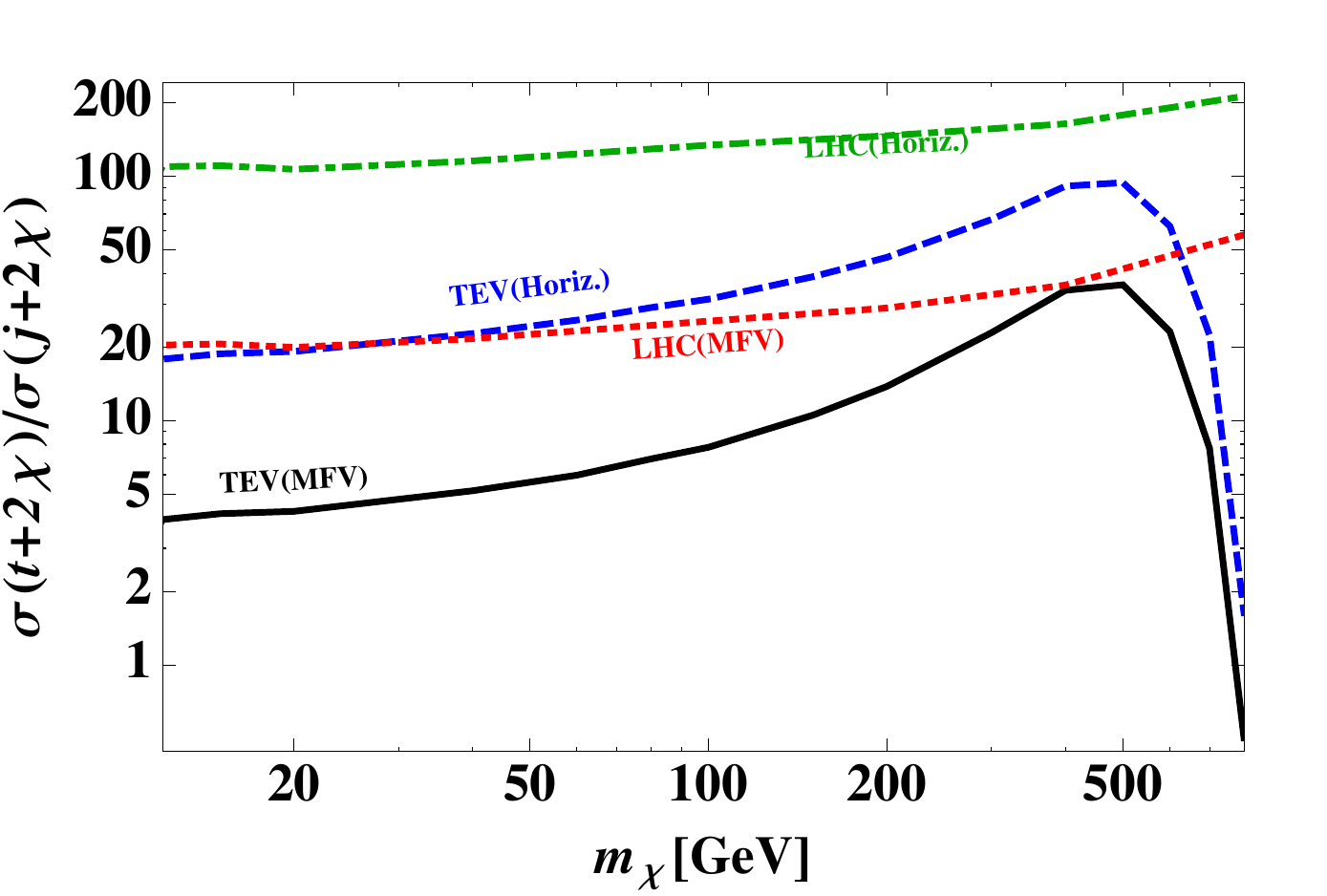}
\caption{ The ratio $\sigma(t+2\chi)/\sigma(j+2\chi)$ as a function of DM mass at Tevatron with $\slashed E_T>80$ GeV (black solid, blue dashed)  and at 7 TeV LHC with $\slashed E_T> 120$ GeV (red dotted, green dot-dashed) for MFV \eqref{MFVestimates} and horizontal \eqref{horizontal:estimates} couplings denoted by (MFV) and (Horiz.), respectively. For quark-DM couplings we assume the EFT limit.}
\label{EFT:plot}
\end{figure}

In the rough estimates \eqref{MFVestimates} we have neglected phase space effects and the role of pdfs. A more quantitative analysis using {\tt MadGraphv4} and CTEQ6L1 pdfs is shown on Fig. \ref{EFT:plot}, where the ratio of production cross sections $\sigma(t+2\chi)/\sigma(j+2\chi)$ as a function of $m_\chi$ is shown for Tevatron and 7 TeV LHC assuming MFV sizes of flavor violating couplings with $b_i=1$ and $y_b=1$. We used $\slashed E_T>80 (120)$ GeV cuts at the partonic level for the Tevatron (LHC) cross sections, following  \cite{Aaltonen:2008hh,Collaboration:2011xw}.  We work in the EFT limit so that the mediator masses drop out in the ratio. The monojet signal is predominantly  produced from charm-gluon initial state resulting in a charm jet in the final state~\cite{Goodman:2010ku}, while in MFV monotop production, the charm-gluon and up-gluon initial state contributions are comparable in magnitude. The monotop signal clearly dominates both at the Tevatron and the LHC.    

{\bf Beyond MFV.}
The above effect is not specific to MFV, and can in fact be much larger for concrete models of flavor.  For instance, in warped extra dimensional models of flavor the coupling of DM to quarks will depend on the localization of the quark zero modes with respect to the zero mode of the mediator. Both large $u_R-t_R$--DM and $c_R-t_R$--DM couplings are possible without violating low energy bounds.  Similarly, the  $u-t$--DM and $c-t$--DM couplings can be enhanced above their MFV estimates in flavor models with abelian or non-abelian horizontal symmetries. 

As an illustration let us assume that the structure of quark Yukawas is due to spontaneously broken horizontal symmetries \cite{Leurer:1992wg}, i.e. that they are generated through a Froggatt-Nielsen type mechanism \cite{Froggatt:1978nt}. The quark fields carry horizontal charges $H(\bar u_R^i),  H(\bar d_R^i), H(Q_L^i)$ so that the Yukawas are given by
\beq
(Y_u)_{ij}\sim \lambda^{|H(\bar u_R^j)+H(Q^i)|}, \quad (Y_d)_{ij}\sim \lambda^{|H(\bar d_R^j)+H(Q^i)|},
\eeq
and we assume that the expansion parameter is $\lambda\simeq \sin \theta_C=0.23$, with $\theta_C$ the Cabibbo mixing angle. 
The quark mass matrices after electroweak symmetry breaking are $(M_{d,u})_{ij}=v (Y_{d,u})_{ij}$, where
we assumed a single Higgs with vacuum expectation  expectation value $v$. An assignment of horizontal charges leading to phenomenologically satisfactory quark masses and CKM matrix, is $H(\{Q_L^1,Q_L^2,Q_L^3;\bar u_R^1,\bar u_R^2,\bar u_R^3; \bar d_R^1,\bar d_R^2,\bar d_R^3\})=\{3,2,0; 3,1,0; 3,2,2\}$ \cite{Leurer:1992wg}. 

The horizontal symmetries then also fix the sizes of DM--quark couplings. Assuming that $J_\DM$ does not carry a horizontal charge (an assumption that we will relax below) the Wilson coefficients are
\beq
\begin{split}
C_2^{ij}\sim \lambda^{|H(\bar u_R^i)-H(\bar u_R^j)|},  \quad C_{4}^{ij}\sim \lambda^{|H(Q_L^i)+H(\bar u_R^j)|},
\end{split} \label{horizontal:supp}
\eeq
or explicitly,
\beq
C_{2}\sim 
\begin{pmatrix}
1 & \lambda^2 &\lambda^3\\
\lambda^2 & 1 &\lambda\\
\lambda^3 & \lambda & 1
\end{pmatrix}, 
\quad
C_{4}\sim 
\begin{pmatrix}
\lambda^6 & \lambda^4 &\lambda^3\\
\lambda^5 & \lambda^3 &\lambda^2\\
\lambda^3 & \lambda & 1
\end{pmatrix}.\label{horizontal:estimates}
\eeq
The constraints from $D-\bar D$ mixing require that the mediator masses are $\Lambda\gtrsim 5$ TeV for $C_2$ (vector case) and $\Lambda\gtrsim 200$ GeV for $C_4$ (scalar mediator). For the case of scalar mediators close to the bound the EFT description is not adequate. The mediators are produced on-shell, a situation that we will cover shortly. Nevertheless, note that the flavor violating couplings in $C_4$ are quite large, $\sim \lambda$ for $\bar c_L t_R \chi^\dagger \chi$, instead of $\sim \lambda^2 y_b^2$ that one would obtain in the MFV counting. The flavor conserving DM production is suppressed compared to flavor violating one. For instance, the partonic cross section for  $c_R g\to c_L +2\chi$ is $(\lambda^2)^2\sim {\mathcal O}(10^{-3})$ suppressed compared to $c_R g\to t_L+ 2\chi$ (see also Fig. \ref{EFT:plot}).

The above hierarchy between flavor violating couplings in $C_2$ and $C_4$ could be changed in other models of flavor, for instance in warped extra dimensional scenarios. It is conceivable that $C_2$ would have large couplings between light and top quark, depending on the profiles of zero modes~\cite{Giudice:2011ak}. 

{\bf Flavorful DM.}
So far we have assumed that DM does not carry a flavor quantum number. Let us next relax this assumption and consider a case where DM carries a nonzero horizontal charge. For simplicity let us assume that DM is a scalar. 
In this case we have two distinct cases for the DM current 
\beq
\J_\DM^{(0)}=\chi^\dagger \chi, \qquad \J_\DM^{(1)}=\chi^2.
\eeq
The current $\J_\DM^{(0)}$ is neutral under horizontal symmetries so that the same analysis as above applies. The second current, $\J_\DM^{(1)}$, on the other hand, carries a nonzero horizontal charge. This can have striking phenomenological implications for the DM production signals at colliders. For instance, if the DM horizontal charge $H(\chi)$ equals $1/2(H(t_L)-H(u_R))$ the $\bar t_L u_R \chi^2$ would have a coupling constant $C_4^{31} \sim O(1)$, with $t+2\chi$ the largest production channel. Note that in this case the flavor violation in the production is only apparent since DM carries away a nonzero horizontal charge.

Another interesting example is DM that is part of a flavor multiplet \cite{Batell:2011tc}. This might be because the underlying flavor symmetry is non-Abelian and $\chi$ is a  part of the flavor multiplet. This can again lead to production of DM through seemingly flavor violating signatures with $t+2\chi$ (one of) the dominant production channels. As a concrete example consider the case of MFV, where DM is in $(3,\bar 3, 1)$ of the flavor $SU(3)_Q\times SU(3)_U\times SU(3)_D$ and the flavor conserving interaction Lagrangian $\epsilon^{ijk} \epsilon^{abc} \bar u_R^i Q_L^a H \chi^{jb} \chi^{kc}$ leads to both $j+\slashed E_T$ and $t+\slashed E_T$ signatures that are unsupressed.

Yet another possibility that can lead to the same type of DM collider signature is a case of composite DM. Let us assume that DM is the lowest lying state of a strongly coupled sector that gets most of its mass from new strong interactions, not from the Yukawa interaction (in the same way as low lying resonances in QCD). In this way one can have an approximately degenerate multiplet of dark states (the lowest being the DM), but each carrying a different horizontal charge despite mass degeneracy.

\begin{figure}
\includegraphics[width=0.45\textwidth]{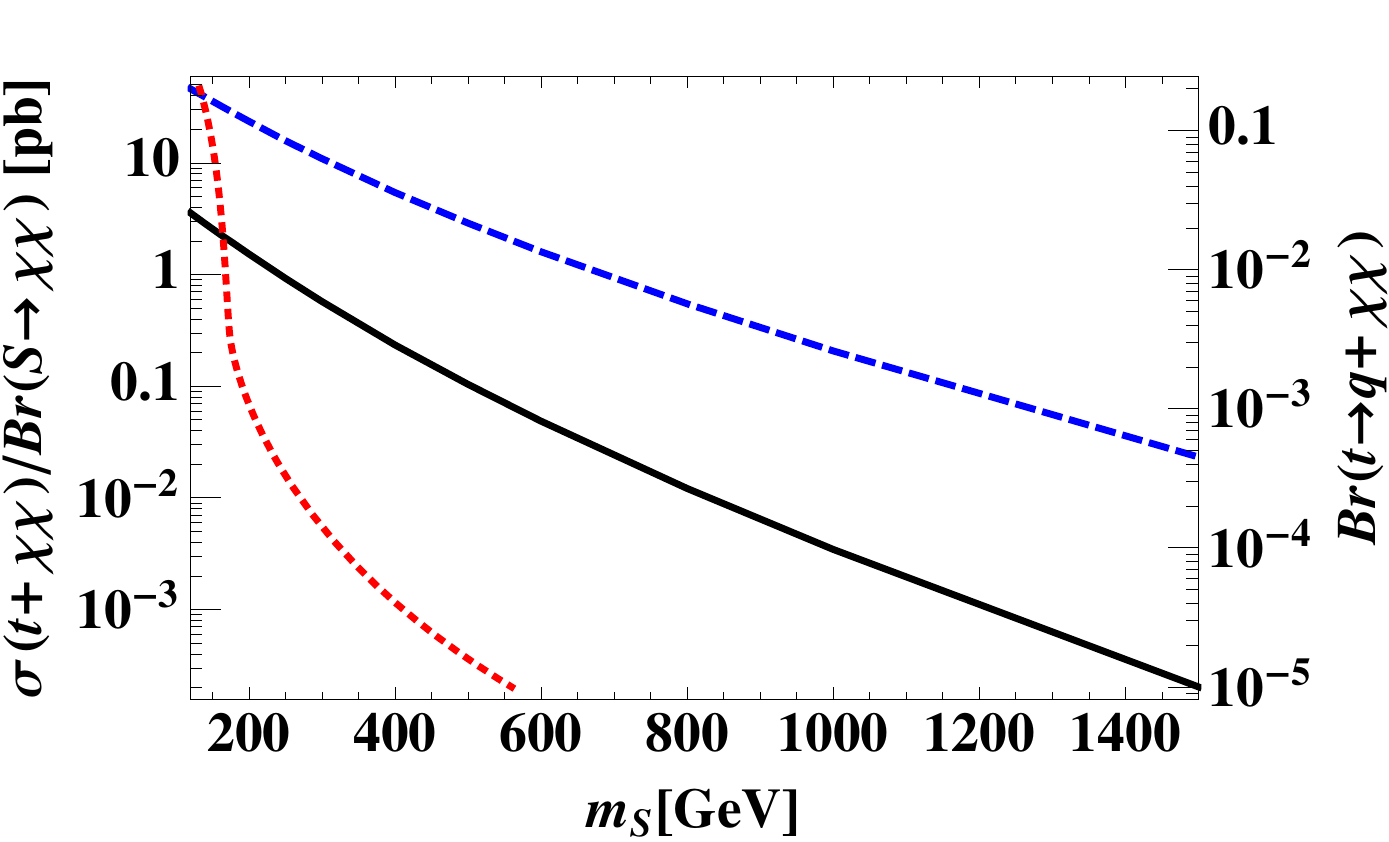}
\caption{The $m_S$ dependence of $Br(t\to j+\chi \chi)$ (red dotted) and of $\sigma(t+2\chi)$ at 7 TeV LHC in model \eqref{Smodel:Lint} for $g_L^u=1$ (blue dashed) and $g_L^c=1$ (black solid), keeping all other $g_i=0$ in each case.}
\label{Smodel:plot}
\end{figure}

{\bf On-shell production of mediators.}
The largest $t+\slashed E_T$ signal can be expected, if the mediators can be produced on-shell. There are two classes of models that can lead to large $t+2\chi$ signals of DM production at colliders, i) models with a $Z_2$ even SM gauge singlet state $S$ (either scalar of vector) coupling to both DM and quarks, and ii) color triplet $Z_2$ odd mediators $\tilde t$ that are scalars (fermions) if DM is a fermion (scalar). Each leads to a different topology, shown on Fig. \ref{production:diagram} (if $\tilde t$ are $Z_2$ even and $\chi$ carries baryon number, also a topology with $s$-channel resonant production is possible \cite{Andrea:2011ws}). If the mediators are light enough to be produced on-shell, the cross section for $t+2\chi$ will be phase space enhanced compared to our EFT discussion so far, where we had a three-body final state to start with.

\begin{figure}[t]
\includegraphics[width=0.45\textwidth]{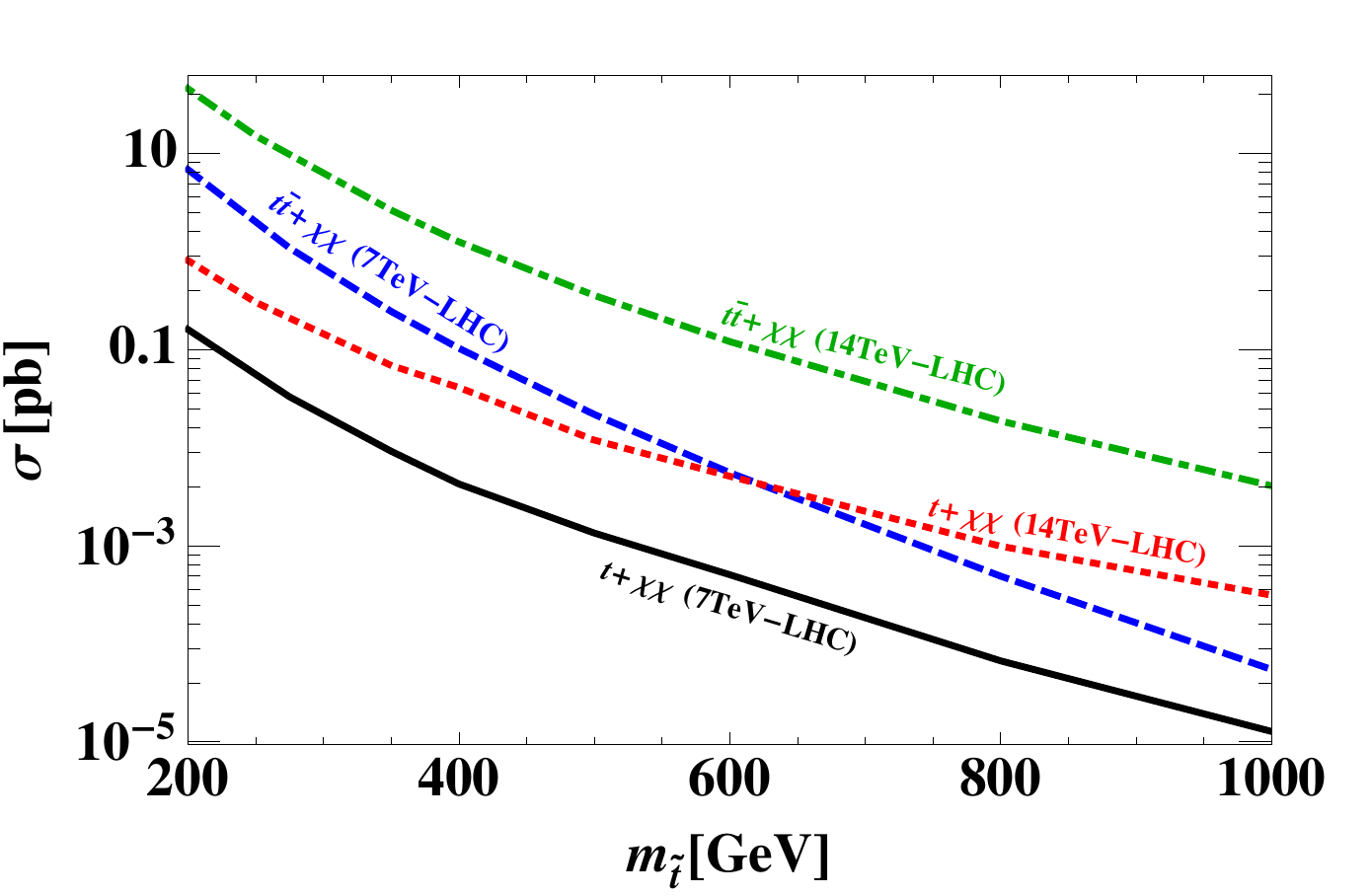}
\caption{Cross sections for single and pair produced $\tilde t_1$ \eqref{stop:Lagr} taking $g_L^{c(u)}=\lambda^{1(3)}$ resulting in $t+\slashed E_T$ and $t\bar t+\slashed E_D$ signal at 7 TeV and 14 TeV LHC.}
\label{stop:plot}
\end{figure}

For illustration we present a toy model example from each of the two classes. First let us consider the case where $S$ and $\chi$ are both scalars, and $S$ has the SM gauge quantum numbers of a Higgs. A model of this sort was considered in~\cite{Li:2011ja}, where FCNC decays of the top were discussed. The relevant part of the interaction Lagrangian after electroweak symmetry breaking (EWSB) is
\beq
\begin{split}
{\cal L}_{\rm int}=g_L^u \bar u_R t_L S  +g_L^c \bar c_R t_L S+ g_R^u \bar t_R u_L S  \\
+g_R^c \bar t_R c_L S+ \lambda v S \chi \chi +h.c.,
\end{split}\label{Smodel:Lint}
\eeq
where the last term arises from $SH^\dagger \chi^2$, and we are intermittently using $S$ for the weak doublet field and its neutral component. On Fig. \ref{Smodel:plot} we show the $t+\chi\chi$ production cross section for two cases, $g_L^u=1$ and $g_L^c=1$, while all the other couplings are taken to zero in each case and $\chi$ is taken massless for simplicity. The results are easily rescaled for the discussed flavor models. With the horizontal charge assignments in \eqref{horizontal:supp}, we would have $g_L^u\sim \lambda^3, g_L^c\sim \lambda, g_R^u\sim \lambda^3, g_R^c\sim \lambda^2$. 
Taking $g_L^c=\lambda=0.23$ the production of top in association with DM can be discovered at the 7 TeV LHC. Using the results of  \cite{Andrea:2011ws} the significance would be $S/\sqrt{S+B}\sim 5,3$ for $m_S=200,400$ GeV with $5$fb${}^{-1}$. Since the irreducible background $3j+Z(\to \nu \bar \nu)$ can be well understood from leptonic $Z$ decays, further improvements with increased statistics can be expected. Note that the for light $\chi$ also $t\to j+2\chi$ decays can be used with the expected 14 TeV LHC reach of $\sim {\mathcal O}(10^{-4})$, however, for increasing $m_S$ monotop signal quickly becomes favored.

A toy example from the second class of models has a $Z_2$ odd majorana fermion $h$, with SM gauge quantum numbers of the Higgs, and two $Z_2$ odd color triplet scalars $\tilde t_{R,L}$ with gauge quantum numbers of right-handed and left-handed  up-quarks. The neutral component of $h$ is DM $\chi$. After EWSB the relevant part of the interaction Lagrangian is 
\beq
{\cal L}_{\rm int}=g_L^u \bar \chi u_R \tilde t_1^*+g_L^c \bar \chi c_R \tilde t_1^*+g_L^t \bar \chi t_R \tilde t_1^*+(L\to R)+h.c,
\label{stop:Lagr}
\eeq
where $\tilde  t_{R,L}$ mix into mass eigenstates $\tilde t_{1,2}$ after EWSB, and we only keep the lowest lying state for simplicity. An example of this model is the Minimal Supersymmetric Standard Model (MSSM) where we only keep the lightest stop and a neutralino which needs to have a large Higgsino component (the impact of flavor violation on neutralino DM within the MSSM has recently been discussed in~\cite{Herrmann:2011xe}). An alternative model realization with a SM singlet DM leading to the same interaction Lagrangian has recently been shown to produce a large forward-backward asymmetry in $t\bar t$ pair production at the Tevatron~\cite{Isidori:2011dp}. Since $\tilde t_1$ is colored, it can be pair produced, leading to $t\bar t+2\chi$ signal. Taking $g_L^t=1, g_L^c=\lambda, g_L^u=\lambda^3$ we compare on Fig. \ref{stop:plot} the $t+2\chi$ and $t\bar t+2\chi$ cross sections at the 7 TeV and 14 TeV LHC as a function of $m_{\tilde t}$, taking $\chi$ again massless for simplicity. For this choice of parameters pair production yields an order of magnitude larger signals. This hierarchy can change if $g_L^{c,u}$ are larger in reality, or if $Br(\tilde t_1\to t\chi)<100\%$ since the pair production $t\bar t+2\chi$ signal scales as this branching ratio squared. In either case the cross sections are large enough that a discovery is possible at the LHC with increased statistics.

{\bf Conclusions.} We have shown that a novel $t+\slashed E_T$ signature is an interesting search channel for DM production at the LHC, where with reasonable size of flavor violation the discovery can be made already at the 7 TeV LHC with a few fb${}^{-1}$ of data. For light DM,  $t\to j+ \slashed E_T$ decays offer another interesting search mode.

This work is supported in part by the European Commission RTN  network, Contract No. MRTN-CT-2006-035482 (FLAVIAnet) and by the Slovenian Research Agency. 

\end{document}